\documentstyle[aps,prl,epsf]{revtex}

\begin{document}

\twocolumn[\hsize\textwidth\columnwidth\hsize\csname
@twocolumnfalse\endcsname

\draft
\date{\today}
\title{{\bf Magnetic Raman Scattering of Insulating Cuprates}}
\author{J.M.Eroles, C.D.Batista, S.B. Bacci and E.R. Gagliano}
\address{Comisi\'on Nacional de Energ{\'\i}a At\'omica}
\address{Centro At\'omico Bariloche e Instituto Balseiro}
\address{8400 S.C. de Bariloche (RN), Argentina.}
\maketitle

\begin{abstract}
We study the $B_{1g}$ and $A_{1g}$ Raman profiles of M$_{2}$Cu$O_{4}$ ( with
M= La, Pr, Nd, Sm, Gd), Bi$_{2}$Sr$_{2}$Ca$_{0.5}$Y$_{0.5}$Cu$_{2}$O$_{8+y}$%
,  YBa$_{2}$Cu$_{3}$O$_{6.2}$ and PrBa$_{2}$Cu$_{2.7}$Al$_{0.3}$O$_{7}$
insulating cuprates within the Loudon-Fleury theory, in the framework of an
extended Hubbard model for moderate on-site Coulomb interaction $U$. We
calculate the non-resonant contribution to these Raman profiles by using
exact diagonalization techniques and analyze two types of contributing
mechanisms to the line shapes: 4-spin cyclic exchange and spin-phonon
interactions. Although these interactions contribute to different parts of
the spectra, together, they account for the enhanced linewidth and asymmetry
of the $B_{1g}$ mode, as well as the non-negligible intensity of the $A_{1g}$
Raman line observed in these materials.
\end{abstract}

\vskip2pc] 
\narrowtext

\section{\bf INTRODUCTION }

Inelastic Raman and neutron scattering experiments on insulating cuprates
provide information about the spin dynamic of the CuO$_{2}$ planes of high-$%
T_c$ precursors \cite{chakra}. The observed short and long wave-length
low-energy excitations have been described using the two-dimensional spin-$%
\frac{1}{{2}}$ antiferromagnetic Heisenberg model (AFH), and the standard
theory of Raman scattering based on the Loudon-Fleury \cite{LF} (LF)
coupling between the light and the spin system. Within this theoretical
scheme, the value of the Cu-Cu exchange constant $J$ has been estimated from
the first moment $M_{1}$ of the $B_{1g}$ Raman line and, in virtual
agreement, from neutron scattering measurements of the spin velocity. For
insulating materials having the Cu atom on different chemical environments,
the estimated value of $J$ results in the range $\sim {(0.10-0.14)}~{eV}$.

Within the AFH framework, numerical and analytical calculations have been
able to describe, the temperature dependence of the spin-spin correlation
length obtained from neutron scattering data \cite{makivic,sigma}, the
temperature dependence of the spin lattice relaxation rate $1/T_{1}$
measured by Nuclear Magnetic Resonance \cite{sokol,imai}, and a prominent
structure, ascribed to 2-magnon scattering, of the $B_{1g}$ Raman line \cite
{rajiv,nos,manousakis2,girvin,sandvik}. Although, all these theoretical results
suggest the AFH on the square lattice as a {\it very good starting point}
for describing the spin dynamic of the insulating cuprates, some features of
the Raman lines remain to be explained within the LF theory. In particular,
it is not yet understood the AFH-LF failure to describe the $B_{1g}$
spectral shape and its enhanced width. And of course, the AFH-LF can not
reproduce a non-vanishing $A_{1g}$ response.

The shape of the $B_{1g}$ line presents some characteristic features common
to several insulating cuprates. In fact, at room temperature, a {\it very
similar} line shape has been observed for all members of the M$_{2}$CuO$_{4}$
series (M= {La, Pr, Nd, Sm, Gd} ), Bi$_{2}$Sr$_{2}$Ca$_{0.5}$Y$_{0.5}$Cu$%
_{2} $O$_{8+y}$(BSCYCO), YBa$_{2}$Cu$_{3}$O$_{6.2}$ and PrBa$_{2}$Cu$_{2.7}$%
Al$_{0.3}$O$_{7}$ (PRBACUALO)\cite{sugai,sule1,tomeno,sule2,merkt}. In all
cases, the line extends up to $\sim 8J$, which in the Ising limit, can be
ascribed to multimagnon excitations involving 16 spins. Unfortunately,
calculations on the AFH show a negligible contribution of multimagnon
processes not only for the Raman line but also for the spin structure factor
at the antiferromagnetic wave vector. Another feature is a very intensive
peak located at $\omega \sim 3J$, identified as $2-magnon$ scattering\cite
{rajiv,nos,manousakis2,girvin}. Now, it is believed that the asymmetry of
the line originates on a second hidden peak on the high energy side of the
spectrum at $\omega \sim 4J$ whose intensity is $\sim 25\%$ smaller than the
intensity of the 2-magnon peak \cite{tomeno}. In the Ising limit, the energy
of these magnetic excitations corresponds to 4-magnon scattering. For the
AFH model, they give a very small contribution to the intensity of the
spectrum. On other geometries interesting features also appear. The $A_{1g}$
line has a maximum at a slightly high frequency with respect to the $2-magnon
$ peak\cite{lyons}, while for the $A_{2g}$ and $B_{2g}$ symmetries the
center of the spectrum is located at $\omega \sim 5J$ and $\omega \sim 4J$
respectively \cite{sule3}. Finally, it has been found a slight temperature
dependence of the scattering intensities of the $B_{1g}$ line in the
insulating cuprates. In fact, the rise in temperature from 30 to 273K causes
the $2-magnon$ peak intensity to decrease by only $\sim 10\%$, in contrast
to $\sim50\%$ for the $S=1$ system La$_{2}$NiO$_{4}$ \cite{sugai}.
Consistently, previous calculations show a weak temperature dependence of
the $B_{1g}$ spectrum for $T<J$\cite{nos0} for the spin $\frac{1}{2}$ AFH
model, which in turns suggest the importance of short-range spin-spin
correlations.


In the Fleury-Loudon theory, the state emerging from the application
of the current operator on the ground state is considered as an eigenstate.
In this intermediate eigenstate, the electron and the hole, produced by the photon 
scattering, are very close in real space,creating a charge transfer exiton, 
and the individual propagation of each
particle is neglected. This approximation could be appropriated for the case of the 
cuprates due to the presence of an electron-electron interaction ($U_{pd}$) 
between nearest neighbor sites (oxygen and copper sites). This interaction, not 
included in the Hubbard Hamiltonian, is important to describe the charge excitations
of the cuprates and induces an attractive interaction between the hole and the 
electron \cite{elina}. Due to this interaction, the hole and the electron are close 
in space and there is not free individual propagation of them \cite{wang}. 
Furthermore, the disorder present in the cuprates \cite{choten} tends to localize
the electron-hole pairs.  Chubukov and Frenkel tried to include the resonant 
effects in the frame of the one band Hubbard model \cite{chubukov-frenkel}. 
Sch\"onfeld {\it et al.} \cite{schonfeld}
calculated the resonant contributions but they could not properly describe
the $B_{1g}$ profile, and the observed $A_{1g}$ line remains to be
explained. In both cases, they did not include the effect of the $U_{pd}$ 
interaction to describe the charge excitations and considered that the electron 
and the hole propagate individually. 
For the reasons given above, we think that the Fleury-Loudon
theory is an appropriated framework to describe the Raman spectra of the 
cuprates.

Even when the resonant effects may be needed
to explain the dependence of the Raman spectra on the incoming laser
frequency, the main features of the line studied in our work persist over the wide range of
frequencies used in the experiments \cite{sugai,sule1,merkt,lyons,ohana,blumberg} and are
essential the same than out of resonance. This point is shown in, e.g., the
recent work in PrBa$_{2}$Cu$_{2.7}$Al$_{0.3}$\cite{merkt} where the Raman
spectra out of resonance was measured and the main features of the $B_{1g}$ line
shape remain unaltered. 
Besides, the LF theory has been the standard framework to describe the Raman spectra
of other antiferromagnetic $S=1/2$ compounds \cite{singh,gros}. So, it is an
interesting issue whether the persistent aspects of the Raman line can be
well described within the LF theory with an appropriate model. 

Previous theoretical work using LF theory tackled the problem of the line's
shape by using series expansions, variational Monte Carlo, interacting
spin-waves and exact diagonalization techniques on small clusters \cite
{rajiv,nos,manousakis2,girvin,sandvik}. 

Different theoretical scenarios have been proposed for describing the
anomalously enhanced width of the $B_{1g}$ line, and although at a first
sight, quantum fluctuations could be the main contribution to the width of
the $B_{1g}$ line, this last hypothesis has been rejected after measurements
on the $S=1$ system NiPS$_{3}$ for which a relative width comparable to the
insulating cuprates has been observed \cite{merlin}. A different proposal
has been motivated by the temperature dependence of the 2-magnon peak,
namely, an increase of the Raman linewidth with increasing temperatures \cite
{knoll}. This has been considered as a strong indication of a phonon
mechanism for the line's broadening and its effect has been analyzed through
a non-uniform renormalization of the exchange constant $J$ \cite{nori}.
Another argument which support the phonon mechanism is that although the
second cumulant $M_{2}$ of the $B_{1g}$ Raman spectrum is almost $J$-
independent for the materials mentioned above, its exchange constant changes
by $\sim 20\%$. This indicates a linewidth dominated by non-magnetic
contributions such as intrinsic disorder of the spin lattice, defects or
phonons. Of course, this does not rule out other magnetic mechanism
providing scattering at frequencies higher than the 2-magnon peak.

So far, the phonon mechanism alone is insufficient to describe the line
asymmetry due to important contributions of 4-magnon scattering \cite{knoll}%
. This requires to go beyond the minimal AFH model. An effective description
of the insulating cuprates based on the single-band Hubbard model \cite{lema}%
, recently appeared, gives a theoretical framework to go beyond the AFH
model and provides additional multispin interactions which finally could
contribute not only to the Raman spectra but also to the midinfrared
phonon-assisted optical conductivity\cite{lorenz}. In this work, we extend
previous calculations of the $B_{1g}$ and $A_{1g}$ lines based on the
minimal AFH model by including multispin and spin-phonon interactions. We
calculate the Raman spectra on finite-size systems using the now standard
Lanczos method\cite{lanczos} for the series of materials mentioned above.
Since inelastic light-scattering measurements probe short-wavelength spin
fluctuations, we expect finite size effects to be small \cite{nos1}. We find
that two mechanisms, multispin and spin-phonon interactions, contribute to
the width and shape of the $B_{1g}$ Raman line, and at the same time, allow
an otherwise forbidden $A_{1g}$ response.

\section{\bf RAMAN SPECTRA AND EFFECTIVE SPIN MODEL }

The scattering of light from insulating antiferromagnets at energies smaller
than the charge-transfer gap $\sim (1.5-2)eV$ can be described
phenomenologically, by introducing a Raman operator, based on the symmetries
of the magnetic problem. For the one-dimensional irreducible representations
of the square, these operators are given by

\[
O_{B_{1g}} = \sum_{{\bf i}} {\vec S_{{\bf i}}} . {( {\vec S_{{\bf i}+{\bf %
e_{x}} }} - {\vec S_{{\bf i}+{\bf e_{y}}}} )}. 
\]

\[
O_{A_{1g}} = \sum_{{\bf i}} {\vec S_{{\bf i}}} . {( {\vec S_{{\bf i}+{\bf %
e_{x}} }} + {\vec S_{{\bf i}+{\bf e_{y}}}} )}. 
\]

\[
O_{B_{2g}} = \sum_{{\bf i}} {(\vec S_{{\bf i}}} . {\vec S_{{\bf {%
i+e_{x}+e_{y}}} }} - {\vec S_{{\bf i}+{\bf e_{x}}}}. {{\vec S_{{\bf i}+{\bf %
e_{y}}}})} 
\]

\[
O_{A_{2g}} = \sum_{{\bf i}}{\epsilon_{\mu \nu}} {\vec S_{{\bf i}}} . {({\vec %
S_{{\bf i}+{\bf e_{\mu}}}}\times {\ {\vec S_{{\bf i}+{\bf e_{\nu}}}})}}. 
\]

\noindent where $\mu = \pm x, \pm y$, and $\epsilon_{\mu \nu} =
-\epsilon_{\nu \mu}= -\epsilon_{-\mu \nu}$. Here ${\bf e_{x}}$ and ${\bf %
e_{y}}$ denote unit vectors in $x$,$y$ directions of a square lattice. The
Raman Hamiltonian is proportional to one of these operators, being the
prefactor matrix elements of the dipolar moment and therefore depend on
microscopic details of the system. Here, we will assume all them equal to
one and therefore we will be unable to compare the relative intensities of
spectra of different symmetries. Microscopically, these Raman operators can
be derived from the theory of Raman scattering in Mott-Hubbard systems\cite
{shastry}. In the strong-coupling limit of the Hubbard model, they appear
quite naturally as the leading-order contributions in an expansion in $%
\Delta = t/(U-\omega)$. Of course, for the $2D$ AFH model, the $O_{A_{1g}}$
operator commutes with the Hamiltonian, so the $A_{1g}$ response vanishes at
the lowest order and terms of order $\Delta^3$ are required to have a small
but finite response.

The Raman spectrum $R_{\Gamma}(\omega)$ at $T=0$, is given by

\[
R_{\Gamma}(\omega)=\sum_{n} {|\langle\phi_{0}|O_{\Gamma}|\phi_{n}%
\rangle|^{2} } \delta(\omega-E_{n}+E_{0}) 
\]

\noindent where $\Gamma$ is $A_{1g},B_{1g},A_{2g}$ or $B_{2g}$, $|\phi_{0}>$
is the ground-state, $E_0$ is the ground-state energy, and $|\phi_{n}>$ is
an excited state of the system with energy $E_{n}$. The spectrum $%
R_{\Gamma}(\omega)$ can be obtained from a continued fraction expansion of
the diagonal matrix element of the resolvent operator $1/(\omega+E_{0}+i%
\delta-H)$ between the state $O_{\Gamma}|\phi_{0}>$. Here $\delta$ is a
small imaginary part added to move the poles away from the real axis \cite
{lanczos}.

Let us describe first the $B_{1g}$ Raman spectrum of the $2D$
antiferromagnetic Heisenberg model.
It has mainly two contributions: (a) a
very strong {\it two-magnon} structure located at $\omega \sim 3J$ and (b) weaker 
{\it 4-magnon} processes at higher-energies. 
In a recent work, A. Sandvik {\it et al.}
 \cite{sandvik} calculated numerically
the Heisenberg $B_{1g}$ spectrum for lattices up to 6$\times6$ sites. They found a 
change in the strong structure at $\omega \sim 3J$ in the
exact spectrum going from 4$\times$4, with a single two magnon dominant
peak, to 6$\times$6, where two equal sized peaks appear. Unfortunately this is 
the bigger system available nowadays and it is not clear if these two peaks
are not an spurious characteristic of this cluster since this kind of structure is not
observed in bigger systems calculated with Monte Carlo-Max Entropy\cite{sandvik}. 
They concluded that the Heisenberg model can not describe the broad $B_{1g}$
spectrum and of course it gives a zero $A_{1g}$ profile even for that size of clusters.
The $4-spin$ excitations are
generated by flipping two pairs of spins of the Neel background in a
plaquete or in a column with an energy cost of $4J$ and $5J$ respectively.
The intensity of these high-frequency processes is {\it very small} $<10\%$%
\cite{nos,nos1}, and cannot account for the asymmetry observed for this mode 
\cite{nota1} on the high frequency side of the spectrum. Since a distinct
shoulder appears at $\omega \sim 4J$, it has been proposed that the $%
4-magnon $ channel is enhanced by 4-spin cyclic exchange interactions \cite
{sugai}. In fact, finite cluster calculations \cite{kuramoto,roger,nos2} for the
undoped compound have shown that although within the CuO$_{2}$ planes the
leading term in the effective spin Hamiltonian is the antiferromagnetic
Heisenberg exchange interaction, other terms are required to describe the
low energy spin excitations. These higher-order terms appear naturally by
performing a canonical transformation up to fourth order on the Hubbard model \cite{mcdonal}. A
single-band Hubbard model can simultaneously describe the low-energy charge
and spin responses of insulating Sr$_2$CuO$_2$Cl$_2$ \cite{lema}. Hence, our
starting model to describe the spin dynamic of all the compounds mentioned
before, is an extended Hubbard Hamiltonian, including next-nearest-neighbor
(NNN) hopping $t^{\prime }$. By means of a canonical transformation to
fourth order in $t$ and second order in $t^{\prime }$, the effective spin
Hamiltonian is written as follows,

\begin{eqnarray*}
H_{eff} &=&J\sum_{i,\delta }\left( {\bf S}_{i}{\bf S}_{i+\delta }- \frac{1}{4%
}\right) + J^{\prime }\sum_{i,\delta ^{\prime }}\left( {\bf S}_{i}{\bf S}
_{i+\delta ^{\prime }}-\frac{1}{4}\right)  \\
&&+J^{\prime \prime }\sum_{i,\delta ^{\prime \prime }}\left( {\bf S}_{i}{\bf S}%
_{i+\delta ^{\prime }}-\frac{1}{4}\right)  \\ 
&&+K\sum_{\left\langle i,j,k,l\right\rangle } ({\bf S}_{i}{\bf S}_{j})({\bf S}%
_{k}{\bf S}_{l}) + ({\bf S}_{i}{\bf S}_{l})({\bf S}_{j}{\bf S}_{k}) \\
&& - ({\bf S}_{i}{\bf S}_{k})({\bf S}_{j}{\bf S}_{l})
\end{eqnarray*}

\noindent with $J=4t^{2}/U-64t^{4}/U^{3},\;J^{\prime }=4t^{\prime
2}/U+4t^{4}/U^{3},\;J^{\prime \prime}=4t^{4}/U^{3},\; $and $K=80t^{4}/U^{3}.$
$\delta $ $\delta ^{\prime }$ $\delta ^{\prime \prime }$
runs over NN; 2nd NN; 3th NN and $\left\langle i,j,k,l\right\rangle $ means the
sum over groups of four spins in a unit square. The effect of the cyclic
exchange interaction on the antiferromagnetic ground state of the $2D$
Heisenberg model has been studied showing that for small values of $K/J$,
the antiferromagnetic phase is stable\cite{chubukov}. The staggered
magnetization increases as $K/J$ is increased, has a maximum at $\sim 0.75$,
and then decreases showing a transition to a spin canted region. In
particular, for the value of $K$ estimated for the cuprates, the system is
still in the antiferromagnetic phase and its staggered magnetization is $%
m_{st}\sim 0.58$. The reduction of the three-band model onto the single band
Hubbard model for realistic values of the multiband parameters, indicates
that the effective hopping $t$ is bounded between 0.3eV and 0.5eV while $%
U/t\sim 7-10$ \cite{3A}. Furthermore, the derivation of the one band Hubbard
Hamiltonian given by Sim\'on and Aligia ( see Ref.\cite{3A} ) for the
parameters obtained from LDA calculations for La$_2$CuO$_4$, gives $t\sim
0.45eV, U/t\sim 7.6, t^{\prime}/t\sim 0.15$. Deviations from these values
are expected for different materials, in particular, for the values of the
long-range hoppings. In Table I, we show the values of the parameters for
the 7 systems studied in this work.

\section{\bf RESULTS }

Let us examine first the role of multispin interactions on the spectra.
Figure 1 shows the calculated Raman spectrum at different symmetries for $%
U=9.5t$ and $t^{\prime }=0.28t$. The $B_{1g}$ line shows the same features
as for the Heisenberg model, being the main difference the gain of spectral
weight at high-frequencies, Fig.1(a). The $4-magnon$ processes are now
enhanced having a relative intensity to the $2-magnon$ peak of $\sim 20\%$.
This is in agreement with Ref.\cite{tomeno}, where it has been found that
the deviation from the symmetric part of the $B_{1g}$ mode is mainly due to
a peak located at $\omega \sim 4J$ whose intensity is about $20\%$ of the
2-magnon peak. For the $A_{2g}$ and $B_{2g}$ modes, Fig.1(c), the center of
gravity of the spectra agree rather well with the experimental data\cite
{sule3}.

\begin{figure}
\narrowtext
\epsfxsize=4.5truein
\epsfysize=3.3truein
\vbox{\hskip 1.truein
\epsffile{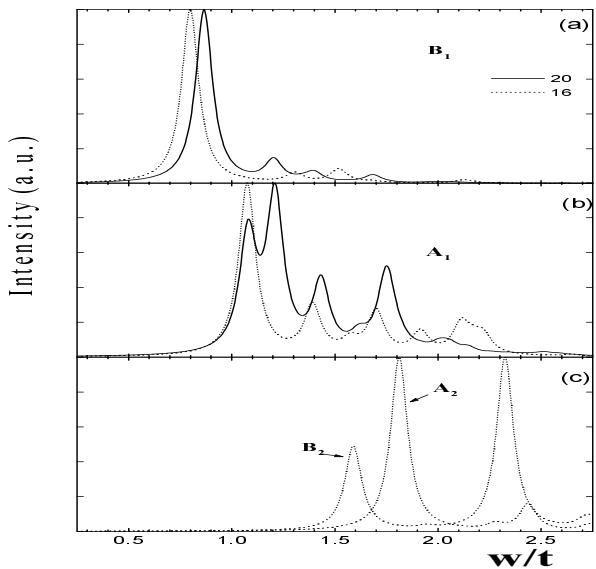}}
\medskip
\caption{ $B_{1g}$ and $A_{1g}$ Raman spectra for 
20 and 16 clusters. Calculations were done
for the effective spin model with no spin-phonon interaction.
The microscopic parameters corresponds to {\protect YBa$_{2}$Cu$_{3}$O$_{6.2}$ }
( see Table I ).}
\end{figure}

However, calculations for the $A_{1g}$ line, Fig.1(b), shows that the
maximum is shifted by $\sim J$ respect to the two magnon peak, in
disagreement with the experimental shift. Although the 4-spin cyclic
exchange interaction provides a mechanism for some features of the $B_{1g}$
mode and yields responses in the correct frequency range for the $A_{2g}$
and $B_{2g}$ lines, it does not give correctly the frequency range for the $%
A_{1g}$ spectrum neither the full linewidth of the $B_{1g}$ line. As was
mentioned before, this can be due to sample inhomogeneities, additional
strong interactions with the spin excitations and/or disorder. Among all of
them, the spin-phonon interaction seem to be the best candidate.

The occurrence of a strong Peierls-type Fermi-surface instability involving
breathing-type displacements of the oxygen atoms of the CuO$_2$ planes
suggests that spin-phonon interactions could contribute to the damping of
the spin excitations of the insulating compounds. The in-plane phonon modes
have small energy compared to the exchange interaction. Indeed, infrared
data for La$_{2}$CuO$_{4}$ gives frequencies of the O stretch modes as $550$
and $690~cm^{-1}$. This corresponds to a bond-stretching force constant $K$
of $\sim 7.5$ and $\sim 11~eV/ {\AA}^{2}$ respectively \cite{weber} $i.e.%
\sqrt{<x^{2}>} \sim 0.1{\AA}$. The electron-phonon interaction on the real
material, modifies the exchange interaction parameter of the effective spin
Hamiltonian $mainly$ through the dependence of $t_{pd}$ and the
charge-transfer energy on the $Cu-O$ distance $d_{Cu-O}$ \cite{ohta}. In the
spin-wave approximation, the lowest order contribution to the damping of
spin excitations due to the spin-phonon interaction is proportional to $%
|\nabla J|^{2}$ \cite{cottam}. Because along the M$_{2}$CuO$_{4}$ family, $J$
changes almost linearly with $d_{Cu-O}$, $|\nabla J|$ takes the $constant$
value $\sim 4350~cm^{-1}/\AA$, so the maximum contribution to the exchange
constant due to the spin-phonon coupling is $\Delta J \sim \pm 435~cm^{-1}$.

The spin-phonon interaction produces, through quantum fluctuations, some
kind of dynamically induced disorder which can be seen by short-wavelength
spin excitations. While the spin dynamic is driven by $\hbar \omega _{s}=J$,
the phonon movement is by $\hbar \omega _{ph}\sim (20-60)meV~<J$, then the
spin system looks the oxygen atoms as frozen in different positions during $%
each$ spin exchange process. Since the phonon energy is much smaller than
the exchange parameter, as a first approximation, the phonon energy can be
taken as zero, then the spin-phonon interaction shows up through
fluctuations of the phonon field. In this limit phonons can be modeled by
static disorder \cite{Rojo}, in other words, the fluctuations of the phonon
field can be modeled by a random distribution $P(J,D)$ of the exchange
parameter of width $D$ centered around $J$. For the one-band Hubbard model
this correspond to a distribution for hoppings, centered around $%
t\;(t^{\prime })$ with dispersion $Dt\;(Dt^{\prime })$. In what follows we
assume a Gaussian distribution for them. The value of $D$ is of order $%
\lambda \sqrt{<x^{2}>}/t \sim (0.07-0.7)$eV for an electron-phonon coupling $%
\lambda \sim (0.3-3.0)$ \cite{weber}. Since the real value of $D$ depends on
the details of the microscopic model used for the planes as well as the
material itself, we fixed $D=0.13$ which gives the better overall
description of the mentioned insulators.

Fig. 2 shows a comparison between experimental and calculated $A_{1g}\;$and $%
B_{1g}$ spectra for seven different compounds. Table II presents the
calculated and experimental values for the first two cumulants. All the
spectra of Fig.2 results from a quenched averaged over 150 samples. The
calculated $B_{1g}$ response shows many of the experimentally observed
features. Indeed, the characteristic $2-magnon$ peak is located at $\omega
\sim 3J$, the spectra extend up to $8J$ and significant spectral weight is
found at $\omega \sim ~(3.5-4)J$. The $A_{1g}$ scattering appears in the

\begin{figure}
\narrowtext
\epsfxsize=3.7truein
\epsfysize=6.0truein
\vbox{\hskip .5truein
\epsffile{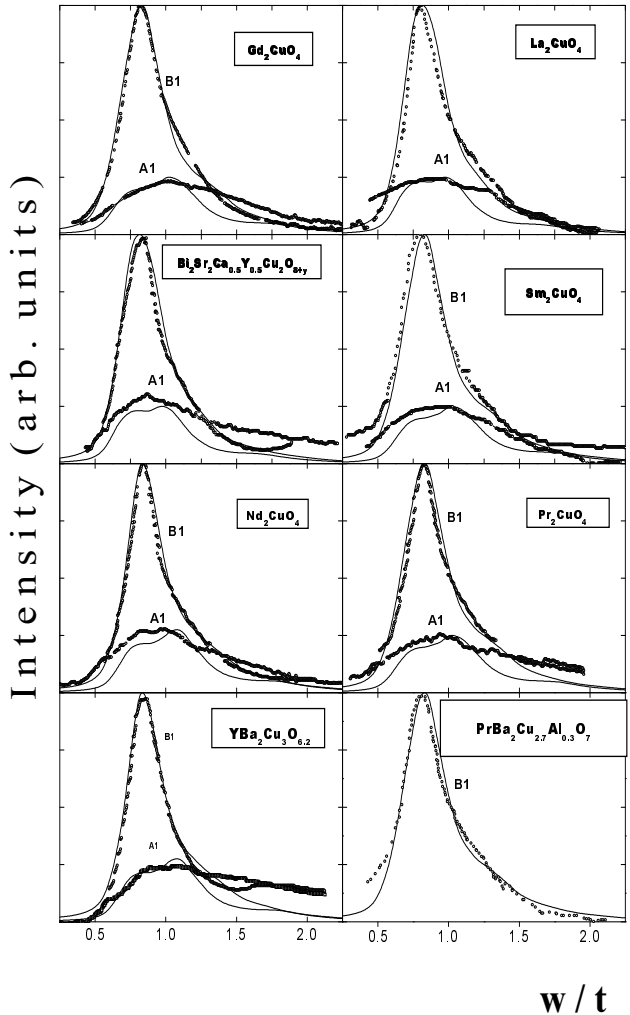}}
\medskip
\caption{ $B_{1g}$ and $A_{1g}$ Raman spectra for the 20
sites cluster. The value of the parameters are given in Table I and $D=0.13$.
Scattered symbols correspond to experimental data taken form 
Ref.{\protect\cite{sugai,sule1,tomeno,sule2}} while solid lines are the result of theoretical 
calculations using the effective spin model taking into account fluctuations
of the phonon field. }
\end{figure}

\noindent same frequency range of the $B_{1g}$ response $2J<\omega <4J$ and the center
of gravity of this line is slightly shifted to the right respect to the $%
2-magnon$ peak. As it was stated in Ref. \cite{tomeno}, the $B_{1g}$ line
can be well fitted with two gaussian curves. One centered at $\omega \sim
3J, $ and the other at $\omega \sim 4J$ with a relative intensity of 25\%
respect to the first one. Hence, from the explanation given above and the
results presented in Ref.\cite{nori}, it is clear that the Heisenberg model,
even with the inclusion of spin-phonon interaction, cannot reproduce
simultaneously the width and the asymmetry of the $B_{1g}$ peak. It is
necessary to take unrealistic values of disorder ($D=0.5J$) \cite{nori} in
order to transfer spectral weight from the $2-magnon$ peak to higher
energies. From Figs.1 and 2, it becomes clear that the inclusion of
additional terms ($J^{\prime }$and $K$) are necessary to reproduce the whole
shape of the $B_{1g}$ line. These terms generate the whole structure of the $%
A_{1g}$ line because $O_{A_{1}}$ does not commute with the Heisenberg
Hamiltonian. This structure has two main contributions of approximately same
weight. The first is the same as the $2-magnon$ $B_{1g}$ peak, now allowed
in this representation due to the broken symmetry introduced by the
fluctuations of the phonon field. Therefore the intensity of this peak
increases with $D$. The second contribution corresponds to $4-magnon$ states
(see Fig.1(b)) mainly introduced by the $J^{\prime }$ and $K$ terms of the
Hamiltonian. Contrary to the $B_{1g}$ contribution, this one does not have a
regular behavior with $D.$ If we apart from $D=0.13$ (the dispersion used to
describe all $B_{1g}$ lines) the two main $A_{1g}$ peaks will not have the
same intensity and the agreement with experiment will pair off. This point
is clearly seen by comparing Fig.2 with Fig.1 of Ref.\cite{nori} where the $%
A_{1g}$ line, using a Heisenberg model plus disorder, is given.

\section{\bf SUMMARY }

Summarizing, we have studied the role of spin-phonon and multispin
interactions on the line shapes of light-scattering experiments within the
Loudon-Fleury theory for the insulating cuprates. For the insulating parent
materials of high-temperature superconductors, by contrast to conventional
antiferromagnets, the energy of short-wavelength spin-excitations is greater
than the characteristic energy of optical phonons so a non-negligible
contribution of the spin-phonon interaction is expected for the damping of
spin-excitations. Good qualitative agreement for the $B_{1g}$ an the $A_{1g}$ was 
found.

As it was noted before, to describe the dependence of the spectrum with
the incident laser, a resonant theory should be used. But, as it is shown in a 
recent experiment on PrBa$_{2}$Cu$_{2.7}$Al$_{0.3}$O$_{7}$\cite{merkt} 
(see figure 2),
resonant effects become weak as the laser frequency is decreased. For low enough 
frequencies, the system is out of resonance. The main characteristics of the $B_{1g}$
line measured in this conditions, the width of the main peak and the shoulder at higher 
energies, remain unaltered. By including spin-phonon interactions in the framework of
the single-band Hubbard model we can reproduce quantitatively these features. 
The $A_{1g}$ spectra are also well described within this theory. Up to our knowledge, this
is the first time in which a theory with realistic parameters can reproduce the $B_{1g}$ and 
$A_{1g}$ measured spectra. Therefore we conclude that the Loudon-Fleury
theory is a good starting point to describe the Raman spectra of the insulating cuprates.

In a recent work, 
a numerical calculation of the Heisenberg model in clusters as big as 
6$\times$6 was performed and two 2-magnon peaks were found. If this is 
confirmed for bigger clusters, note that Monte Carlo-Max Entropy calculations on
bigger clusters do not show such structures, this could lower the
phonon-magnon coupling needed to adjust the experiments.
Unfortunately this is far from nowadays numerical possibilities.

Finite-cluster calculations for 8 insulating cuprates support the view of
two contributing mechanism to the $B_{1g}$ and $A_{1g}$ Raman lines.
Although they contribute to different parts of the spectra, together, they
account for the enhanced linewidth and the asymmetry of the $B_{1g}$ mode,
as well as for a non-negligible response for the $A_{1g}$ Raman line
observed in these materials.

\section{\bf ACKNOWLEDGMENTS }

We would like to thank E.Fradkin, R.M.Martin and M.Klein for useful
discussions at early stages of this work. All of us are supported by the
Consejo Nacional de Investigaciones Cient\'{\i }ficas y T\'{e}cnicas,
Argentina. Partial support from Fundaci\'on Antorchas under grant 13016/1
and from Agencia Nacional de Promoci\'on Cient\'{\i }fica y Tecnol\'ogica
under grant PMT-PICT0005 are gratefully acknowledged.



\twocolumn[\hsize\textwidth\columnwidth\hsize\csname
@twocolumnfalse\endcsname

\newpage

Table I: Values of the single-band parameters for the different insulators.

\begin{tabular}{||l||l|l|l|l|l|l|l|l||}
\hline\hline
& YBa$_{2}$Cu$_{3}$O$_{6.2}$ & $\Pr_{2}$CuO$_{4}$ & Nd$_{2}$CuO$_{4}$ & Sm$%
_{2}$CuO$_{4}$ & La$_{2}$CuO$_{4}$ & Gd$_{2}$CuO$_{4}$ & BSCYCO & PRBACALO
\\ \hline\hline
$U/t$ & 9.5 & 9.5 & 9.5 & 9.5 & 10.5 & 9.5 & 10.5 & 9.5 \\ \cline{1-9}
$t^{\prime }/t$ & 0.28 & 0.30 & 0.28 & 0.30 & 0.30 & 0.30 & 0.30 & 0.30 \\ 
\cline{1-9}
$t$ $\left( eV\right) $ & 0.40 & 0.42 & 0.42 & 0.435 & 0.475 & 0.435 & 0.45
& 0.37 \\ \hline
$J$ $\left( eV\right) $ & 0.138 & 0.145 & 0.145 & 0.150 & 0.154 & 0.150 & 
0.146 & 0.140 \\ \hline
\end{tabular}

\newpage

\vspace{2cm}

Table II: {Values of the cumulants $M_{1}$ and $M_{2}$ for Raman lines $%
B_{1g}$ and $A_{1g}$. First lines correspond to experimental values taken
for Ref.\cite{sugai,sule1,tomeno,sule2}.}

\begin{tabular}{||ll||l|l|l|l|l|l|l|l||}
\hline\hline
&  & YBaCuO & $\Pr_{2}$CuO$_{4}$ & Nd$_{2}$CuO$_{4}$ & Sm$_{2}$CuO$_{4}$ & La%
$_{2}$CuO$_{4}$ & GdCuO & BSCYCO & PRBACAL \\ \hline\hline
& \multicolumn{1}{||l||}{$M_{1}$} & 1.11 & 1.01 & 1.03 & 1.00 & 0.96 & 0.97
& 0.95 & 0.93 \\ 
& \multicolumn{1}{||l||}{} & 1.02 & 0.96 & 1.03 & 1.00 & 0.97 & 0.99 & 0.93
& 0.96 \\ \cline{2-10}
$B_{1}$ & \multicolumn{1}{||l||}{$M_{2}$} & 0.40 & 0.33 & 0.37 & 0.37 & 0.30
& 0.32 & 0.27 & 0.28 \\ 
& \multicolumn{1}{||l||}{} & 0.42 & 0.30 & 0.39 & 0.40 & 0.38 & 0.42 & 0.27
& 0.30 \\ \cline{2-10}
& \multicolumn{1}{||l||}{$M_{2}/M_{1}$} & 0.36 & 0.33 & 0.36 & 0.37 & 0.31 & 
0.32 & 0.29 & 0.30 \\ 
& \multicolumn{1}{||l||}{} & 0.41 & 0.31 & 0.39 & 0.40 & 0.39 & 0.42 & 0.29
& 0.31 \\ \hline\hline
& \multicolumn{1}{||l||}{$M_{1}$} & 1.30 & 1.13 & 1.15 & 1.04 & 1.05 & 1.23
& 1.20 & - \\ 
& \multicolumn{1}{||l||}{} & 1.10 & 1.07 & 1.13 & 1.09 & 1.03 & 1.10 & 1.03
& - \\ \cline{2-10}
$A_{1}$ & \multicolumn{1}{||l||}{$M_{2}$} & 0.42 & 0.39 & 0.45 & 0.37 & 0.38
& 0.42 & 0.46 & - \\ 
& \multicolumn{1}{||l||}{} & 0.35 & 0.33 & 0.37 & 0.36 & 0.33 & 0.37 & 0.35
& - \\ \cline{2-10}
& \multicolumn{1}{||l||}{$M_{2}/M_{1}$} & 0.32 & 0.35 & 0.39 & 0.36 & 0.36 & 
0.34 & 0.39 & - \\ 
& \multicolumn{1}{||l||}{} & 0.32 & 0.30 & 0.33 & 0.33 & 0.32 & 0.33 & 0.34
& - \\ \hline\hline
\end{tabular}

\vskip2pc]

\end{document}